\font\texteusm eusm10 scaled1095
\font\scripteusm eusm7
\font\scriptscripteusm eusm5
\newfam\eusmfam
\textfont\eusmfam\texteusm
\scriptfont\eusmfam\scripteusm
\scriptscriptfont\eusmfam\scriptscripteusm
\def\scri{{\fam\eusmfam I}}
\documentstyle[revtex]{aps}

\begin{document}
\begin{title}
{ On ``hyperboloidal'' Cauchy data for vacuum Einstein equations\\
 and obstructions to smoothness of ``null infinity''}
\end{title}
\author{Lars Andersson\cite{Lars}}
\begin{instit} Department of Mathematics,
Royal Institute of Technology, S--10044 Stockholm, Sweden.
\end{instit}
\author{ Piotr T. Chru\'sciel\cite{Piotr}}
\begin{instit}
Max Planck Institut f\"ur Astrophysik, Karl Schwarzschild Strasse 1,
D--8046 Garching bei M\"unchen, Germany.
\end{instit}
\receipt{October 92, in revised form February 93}

\begin{abstract}
Various works have suggested that the Bondi--Sachs--Penrose decay conditions
on the gravitational field at null infinity are not generally representative
of asymptotically flat space--times. We have made a detailed analysis of the
constraint equations for ``asymptotically hyperboloidal'' initial data and
find that log terms  arise generically in asymptotic expansions. These terms
are absent in the corresponding Bondi--Sachs--Penrose expansions, and can be
related to explicit geometric quantities.
We have nevertheless
shown that there
exists a large class of ``non--generic'' solutions of the constraint equations,
the evolution of which leads to space--times satisfying the
Bondi--Sachs--Penrose
smoothness conditions.
\end{abstract}
\pacs{04.20, 04.30, 04.90}
  \draft
\narrowtext

One of the
milestone predictions of general relativity is that of the
existence of gravitational radiation. A framework for analyzing that aspect
of the theory was proposed by Bondi {\em et al.\/} \cite{BMS} and
Sachs \cite{Sachs}: in
\cite{BMS,Sachs} it was assumed that the metric admits an expansion in terms
of inverse powers of $r$ along lightlike directions, where $r$ is a
luminosity distance ({\em cf.\ }also \cite{Trautman} for some significant early
analysis). It was subsequently realized by Penrose \cite{Penrose} that the
Bondi--Sachs asymptotic
conditions are related to (and in fact equivalent to)
the possibility
of completing the space--time by adding to it a conformal boundary. In the
Penrose approach the hypothesis of the existence of expansions in terms of
$r^{-i}$ is  replaced by
assumptions on the regularity
of the conformally rescaled metric near the conformal boundary. Both in the
Bondi--Sachs and in the Penrose framework the basic open question was, and
still is:
\begin{quote}
Are the Bondi--Sachs--Penrose
asymptotic conditions
compatible with
the behaviour of the gravitational field in a ``sufficiently large'' class of
physical situations?
\end{quote}
It should be stressed that until 1986, those
metrics which were known to
satisfy the appropriate asymptotic conditions ({\em cf. e.g.} \cite{Bi} for a
review), namely the boost-rotation symmetric space--times and the
Robinson--Trautman space--times ({\em cf.\ }also \cite{Schmidt}), were
of a rather special kind.

Significant progress towards the understanding of
this question was made by Friedrich, who showed \cite{Friedrich} that
space--times satisfying the Bondi--Sachs--Penrose conditions could be
constructed provided that sufficiently well behaved initial data ``of
hyperboloidal type'' could be found. The question of the ``largeness'' of the
set of space--times admitting smooth conformal completions was consequently
shifted to the question of the ``largeness'' of the set of initial data
satisfying the conditions spelled out by Friedrich.
The possibility of constructing appropriate initial data was demonstrated in
\cite{ACF} under, however, some rather restrictive hypotheses on the
extrinsic curvature of the Cauchy surface.
In this
letter we present the results of \cite{ACh}
and of \cite{AChWeyl} concerning
the possibility of constructing appropriate initial data using the
conformal Choquet--Bruhat---Lichnerowicz---York \cite{ChY} method.
Our main results are the following:
for generic ``seed fields'' the Cauchy
data constructed by the conformal method starting from a ``smoothly conformally
 compact" Riemannian manifold do {\em not} possess the asymptotic
regularity compatible with Penrose's smoothness requirements for
${\scri}$.
[The notion of non--genericity referred to here is made clear by
equations (\ref{nolog1})--(\ref{nolog2}) below.]
On the other hand, there exists a large class of non--generic
``backgrounds'' for which the solutions display the required regularity.
The ``generic non--smoothness of $\scri$'' suggested by our analysis confirms a
similar observation of Christodoulou and Klainerman \cite{ChKl}. It is
also fully consistent with Winicour's \cite{Winicour} analysis of
Bondi--type
initial data ({\em cf.\ }also \cite{ChS}), as well as with various
approximate
calculations ({\em cf.\ }\cite{Damour} and references therein).

Let us recall the Lichnerowicz--Choquet--Bruhat--York
method for constructing solutions of the constraint equations
(\ref{(A.1)})--(\ref{(A.0)}) of
general relativity, with asymptotic conditions appropriate to the
``hyperboloidal'' initial value problem \cite{Friedrich}, {\em cf.\ }also
\cite{ACF,ACh}. Let ${\bar M}$ be a smooth, connected, Hausdorff, compact
three dimensional
manifold with smooth boundary, set $M = {\rm int}
{\bar M}$, $\partial{\bar M} = \partial M\equiv {\bar M}\setminus M$.
For $0\le k \le \infty$ let $C^k(\bar M)$ denote the space of $k$-times
differentiable tensor fields on $M$ which, in local coordinates near
$\partial M$, extend continuously to $\partial M$  together with all
derivatives of order less than or equal to $k$.
We wish to find initial data $({\tilde g}, {\tilde K})$ which solve
the general relativistic constraint equations
\begin{equation}
R({\tilde g}) = {\tilde g}_{ij} {\tilde g}_{kl} {\tilde
K}^{ik}{\tilde K}^{jl} -
({\tilde g}_{ij} {\tilde K}^{ij})^2\,,\label{(A.1)}
\end{equation}
\begin{equation}
{\tilde D}_{i} ({\tilde K}^{ij} -{\tilde g}_{kl} {\tilde K}^{kl}
{\tilde g}^{ij})= 0\,,\label{(A.0)}
\end{equation}
where ${\tilde D}_{i}$ is the Levi--Civita connection of ${\tilde g}_{ij}$,
and $R(f)$ is the Ricci scalar of a metric $f$, satisfying the condition
\begin{equation}
{\tilde K}\equiv
{\tilde g}_{ij} {\tilde K}^{ij} = 3\,.
\label{(A.2)}
\end{equation}
We may construct such data from a set of
``seed fields'' $(g_{ij}, A^{ij})$, where $g_{ij}$ is any
smooth Riemannian metric on $M$ extending smoothly to $\partial M$,
and $A^{ij}$
is any symmetric, traceless tensor field on ${ M}$
extending smoothly to $\partial M$, via the following procedure:
Let $x$ be any defining
function for $\partial M$, {\em i.e.\ }a function satisfying $x \in
C^\infty({\bar M})$, $x \ge 0$, $x(p) = 0 \Longleftrightarrow p \in \partial
M$,
 and $dx\ne 0$ at $\partial M$. Let $A^{ij} \in C^\infty({\bar M})$ be
symmetric traceless  and   let $X$ be any solution of the equation
\[
D_i\left[x^{-3}\left(D^i X^j + D^j X^i - \frac{2}{3} D_k X^k g^{ij}\right)
\right] \qquad\qquad\qquad\qquad\qquad
\]
\begin{equation}
\qquad \qquad \qquad \qquad \qquad\qquad = -
D_i\left(x^{-2} A^{ij}\right)\,,
\label{(A.3)}
\end{equation}
define
\begin{equation}
L^{ij} \equiv
\frac{\omega^2}{x^{3}} \left(
D^i X^j + D^j X^i - \frac{2}{3} D_k X^k g^{ij}\right) +
\frac{\omega^2}{x^{2}}
A^{ij}\ ,
\label{Ldef}
\end{equation}
where $\omega$
is a solution of the equation
\begin{equation}
\omega \Delta_g \omega - \frac{3}{2} |D \omega|_g^2 + \frac{1}{4}
\omega^2 (R(g) - |L|_g^2) + \frac{3}{2} = 0\,,\label{(A.4)}
\end{equation}
satisfying $\omega \ge 0$, with $\omega x^{-1}$ uniformly bounded on $M$
from above and uniformly bounded away from zero, and where $|L|_g^2 =
g_{ij} g_{kl} L^{ik} L^{jl}$.
Setting
\begin{eqnarray*}
{\tilde g}_{ij} &= &\omega^{-2} g_{ij}\,,\\
{\tilde K}^{ij} &= &\omega^3 L^{ij} +
{\tilde g}^{ij}\,,
\end{eqnarray*}
one obtains a solution of (\ref{(A.1)})--(\ref{(A.0)}) satisfying
(\ref{(A.2)}). Moreover, when $X$ is chosen appropriately
 the fields $({\tilde g}_{ij}, {\tilde K}^{ij})$ satisfy
(in a rough way) the asymptotic conditions appropriate to the ``asymptotically
hyperboloidal'' setting ({\em cf.\ }\cite{Friedrich,ACh}).
The asymptotic behaviour of the initial data as described above is somewhat
reminiscent of that which occurs for 
initial data induced on a
standard hyperboloid in Minkowski space--time, whence the terminology above.

It should be pointed out that equations (\ref{(A.3)})--(\ref{(A.4)})
constitute a nonlinear system of elliptic equations uniformly degenerating at
the boundary $\partial M$, which is at the origin of various difficulties.
Although a vast literature on such problems exists
({\em cf.\/} \cite{literature} and references therein), no
detailed information of the kind needed {\em e.g.\/} for  Friedrich's evolution
theorems \cite{Friedrich} can be found.  

For simplicity we shall
assume throughout this letter
that $\partial M\approx S^2$ --
the two dimensional sphere.
In \cite{ACh} the
following has been shown:
\begin{enumerate}
\item For any $(g, A)$ as above one can find a solution $X $
to (\ref{(A.3)}) such that
\begin{eqnarray*}
\left(\frac{x}{\omega}\right)^2
L^{ij} = &{{U}}^{ij} + x^{2}\log x\,  {U}^{ij}_{\log},
\\
&\qquad {{U}}^{ij},
{U}_{\log}^{ij} \in C^\infty({\bar M})\,.
\end{eqnarray*}
Given any $g$ there exists an open dense set (in the $
C^\infty(\bar M)$ topology) of $A$'s for which
${U}_{\log}^{ij}\big|_{\partial M} \not\equiv 0$
(however, there exists an infinite
dimensional closed subspace of $A$'s for which
${U}_{\log}^{ij}\big|_{\partial M}
\equiv 0$).
If ${U}_{\log}^{ij}\big|_{\partial M}
\equiv 0$, then ${U}_{\log}^{ij}
\equiv 0$ and thus $x^2 \omega^{-2} L^{ij}\in C^\infty({\bar M}) $.
 Let us also note
that in an orthonormal frame
$e_i$ such that $e_A\parallel \partial M$, $A=2,3$,
if we write, in a
neighbourhood of ${\partial M}$,
\[
L^{ij}
= L_0^{ij}(v) + x L_1^{ij}(v) + \cdots\,,
\]
where
$v$ denote coordinates on ${\partial M}$, then we have $L_0^{1i}
\equiv 0$, while both $L_0^{AB}(v) - \frac{1}{2} L_0^{CD} h_{CD} h^{AB}$
and $L_1^{AB}(v) - \frac{1}{2} L_1^{CD} h_{CD} h^{AB}$ are freely
specifiable tensor fields on ${\partial M}$.
$X$ is unique in an appropriate class of functions, 
{\em cf.\/} \cite{ACh} for details.

\item For any $(g, A)$ as above one can find a solution $\omega \in {\cal
A}_{phg}$
of equation (\ref{(A.4)}), where ${\cal A}_{phg}$ denotes the space of
polyhomogeneous
functions on ${\bar M}$:
more precisely, there exists a sequence $\{N_j\}_{j=0}^\infty$ with $N_0 =
N_1 = N_2 =N_3 = 0$, $N_4 = 1$ and functions $\omega_{i,j} \in C^\infty({\bar
M})$ such that
\end{enumerate}
\begin{equation}
\qquad\ 
\omega \sim \sum_{i\ge0} \sum_{j=0}^{N_i} \omega_{i,j} x^i \log^j x
\ ,
\label{omega-expansion}
\end{equation}
\begin{enumerate}
\item[\ ]
where ``$\sim$'' means ``asymptotic to'', in the sense that
$\omega$ minus an appropriately truncated  sum of the form
given by the right hand side of
(\ref{omega-expansion})
vanishes faster than  $x^n$ with any desired $n$,
 and that this property is preserved under differentiation in the
obvious way.
According to  standard terminology,
functions with these properties are called {\em polyhomogeneous},  {\em cf.\
e.g.\/}  \cite{Hormander}.
For an open dense set
of $(g, A)$'s we have $\omega_{4,1}\big|_{\partial M} \not\equiv 0$. If
$\omega_{4,1}\big|_{\partial M} \equiv 0$, then $\omega\in
C^\infty({\bar M}) $.
 \end{enumerate}

Suppose now that one has initial data such that the log terms
described above do not vanish. In such a case the metric will
immediately pick up log terms when time--evolved with Einstein equations, so
that at later times  there will be no decomposition of the three
dimensional metric into a smooth up to the boundary background and a
conformal factor. This shows that it is natural to
consider the above construction under the condition that the seed
fields are polyhomogeneous rather than smooth.
One can show \cite{ACh}
that for any polyhomogeneous Riemannian metric $g$
on $\bar M$
and for any
uniformly  bounded
polyhomogeneous symmetric tensor field $A^{ij}$ on $\bar M$ there exist
solutions $(X,\omega)$ of (\ref{(A.3)}) and (\ref{(A.4)}) such that
$ L^{ij}$ given by (\ref{Ldef})
 is polyhomogeneous and uniformly bounded on
$\bar M$, and
$\omega/x$ is polyhomogeneous, uniformly bounded, and uniformly bounded
away from zero on
$\bar M$.

When $L^{ij} \equiv 0$ and the seed metric is smooth up to the boundary,
the obstructions to smoothness of $\omega$ have been
analyzed in detail in \cite{ACF}. In that reference it has been shown,
in particular, that
$\omega_{4,1}\big|_{\partial M}$ vanishes if
the Weyl tensor of
the unphysical (conformally rescaled) space--time metric is bounded
near ${\partial M}$ ({\em cf.\/} \cite{ACF} for details).
In \cite{AChWeyl} we have extended that analysis to
the case $L^{ij} \not\equiv 0$.
In order to present our results it is useful to define two
tensor fields ${\sigma}^\pm$ defined on the
conformal boundary $\partial M $ of the initial data surface:
\[
{\sigma}^\pm_{AB}
\equiv\bigg(\lambda_{AB} - \frac{h^{CD}\lambda_{CD}}{2} h_{AB}\bigg)
\qquad\qquad\qquad\qquad
\]
\begin{equation}
\label{shear-tensor}
\qquad\qquad
\pm\bigg(K_{AB} - \frac{h^{CD}K_{CD}}{2} h_{AB}\bigg)\bigg|_{\partial M}\,,
\end{equation}
which we shall call the {\em shear tensors of} $\partial
M$. Here  $h_{AB}$ is the induced metric  on $\partial
M$,
$\lambda_{AB}$ is the extrinsic curvature of $\partial
M$ in $(M,g)$, and $K_{ij}$ is the extrinsic curvature of $M$ in the
conformally rescaled space--time metric.

Let us say that a space--time admits a polyhomogeneous $\scri$ if the
conformally
rescaled metric is polyhomogeneous at the conformal boundary; {\em i.e.}, in
local coordinates the components of the conformally rescaled metric are
bounded and polyhomogeneous.
In the case of Cauchy data
constructed as described
above starting from smooth seed fields,
the results of \cite{AChWeyl}
linking the geometry of the boundary of the initial data
surface with
the geometry of the resulting space--time can be summarized as follows:

\begin{enumerate}
\item
Suppose that neither ${\sigma}^+$ nor ${\sigma}^-$ vanishes. Then
{\em  there
exists no development of the initial data with a smooth or
polyhomogeneous $\scri$.}
\item
Suppose that  ${\sigma}^+\equiv 0$ or ${\sigma}^-\equiv 0$; changing the
time orientation if necessary we may without loss of generality assume
that ${\sigma}^+\equiv 0$.
Let $K_{ij}^{\log}$ denote ``the logarithmic part'' of
$K_{ij}$:
\begin{eqnarray*}
K_{ij}=\hat K_{ij}+x^2\log x\,K_{ij}^{\rm log},
\\
\qquad \qquad
\hat K_{ij}\in C^{\infty}(\bar M),\ K_{ij}^{\rm log}\in C^{0}(\bar
M)\cap{\cal A}_{phg}.
\end{eqnarray*}
Then the following holds:
\begin{enumerate}
\item
If  $K_{1A}^{\log}\big|_{\partial M}\not\equiv 0$, then if there
exists a development with a polyhomogeneous $\scri$,
it is essentially polyhomogeneous, {\em i.e.}, no development with
a smooth $\scri$ exists.
[The vanishing of  $K_{1A}^{\log}\big|_{\partial M}$ is actually equivalent
(under the present assumptions) to the vanishing at the conformal boundary of
the Weyl tensor of the
conformally rescaled metric.]
\item
Suppose instead that $K_{1A}^{\log}\big|_{\partial M}\equiv 0$
and  $K_{11}^{\log}\big|_{\partial M}\equiv 0$.
Then
{\em there exists a development which admits a smooth conformal boundary.}
\end{enumerate}
\end{enumerate}

It should be stressed that the results linking the log terms with the
non--vanishing of the Weyl tensor proved in \cite{AChWeyl}
show that the occurrence of shear and
of at least
some of the log terms in
asymptotic expansions of physical fields at $\scri$ is not an artefact
of a bad
choice of a conformal factor, or of a pathological choice of the initial data
hypersurface (within the class of uniformly bounded from
above and uniformly bounded away from zero, locally
$C^2$, conformal factors and $C^1$ deformations of the initial data
hypersurface which
fix $\partial M$): if $\scri$ is {\em not}
shear--free
(by which we mean that
none of the shear tensors $\sigma^\pm$ vanish),
then no conformal transformation will
make it shear free. Similarly if the Weyl tensor does not vanish at $\partial
M$, then no ``gauge transformation'' in the above sense
will make it vanish ({\em cf.\ }\cite{AChWeyl}
for
a more detailed discussion).

The conditions for smoothness--up--to--boundary of an initial data set
can be expressed as {\em local} conditions on the boundary on the seed
fields
$(g_{ij},A^{ij})$.
Let $(x,v^A)$ be a Gauss coordinate system near $\partial M$;
the interesting case is the one in which one of the shear tensors of
the conformal boundary vanishes, which corresponds to the condition
that, changing
$A_{ij}$ to $-A_{ij}$
if necessary,
\FL
\begin{equation}
\bigg(\lambda_{AB} - \frac{\lambda}{2} h_{AB}\bigg)\bigg|_{\partial M}
=
\left(A_{AB} - \frac{1}{2} h^{CD}A_{CD}
 h_{AB}\right)\bigg|_{\partial M}\, .
\label{shear3}
\end{equation}
It can be shown that without loss of generality in the construction of
the initial data
one can assume that
$$
A_{1j}\Big|_{\partial M}=0\ ,
$$
and in what follows we shall assume that this condition holds -- the
equations below would have been somewhat more complicated without this
condition.
Similarly, it is useful to choose a ``conformal gauge''
such that
\[
\lambda\equiv h^{AB}\lambda_{AB}\bigg|_{\partial M}=0\ .
\]
Then the
conditions
for smoothness up to the boundary
of $\omega$ and $L^{ij}$
reduce to
\begin{equation}
\label{nolog1}
 \left [
{\cal D}^A{\cal D}^B\lambda_{AB} + R_{AB}\lambda^{AB} 
  \right ] \Big|_{\partial M} = 0\ ,
\end{equation}
where ${\cal D}$ is the covariant derivative operator of the metric
$h$ induced from $g$ on $\partial M$,
 $R_{ij}$ is the Ricci tensor of $g$, and
\begin{equation}
\label{nolog2}
\left[\partial_xA_{AB}-\frac{1}{2}h^{CD}\partial _x A_{CD} h_{AB}
\right]\Big|_{\partial M} = 0\ .
\end{equation}
Failure of (\ref{nolog1}) or (\ref{nolog2}) will lead to occurrence of
some
log terms in the initial data set.

The overall picture that emerges from the results of
\cite{ACh,ChKl,Winicour,ChS,Damour} and
from the results
described here is that the usual hypotheses of smoothness of $\scri$ are
overly restrictive. These results seem to indicate that
a possible self--consistent setup for an
analysis of the gravitational radiation is that of {\em polyhomogeneous} rather
than smooth functions on the conformally completed manifold, {\em i.e.},
functions that have asymptotic expansions in terms of powers of $x$ and $\log
x$ rather than of $x$ only.
It should, however, be stressed that even though the fact that the physical
fields
$(\tilde g,\tilde K)$ satisfy the constraint equations guarantees the existence
of
some vacuum development $(V,\tilde \gamma)$, it is by no means obvious that
in the case when {\em e.g.\ } ${\sigma}^+\equiv 0$,
the existence of some kind of compactification of $(M,\tilde g,\tilde K)$
implies the existence of some useful conformal completion of $(V,\tilde
\gamma)$.
In the case of
{\em polyhomogeneous hyperboloidal} initial Cauchy data
the question of existence of
conformal completions of $(V,\tilde \gamma)$ with three dimensional
boundaries is the most important unsolved
mathematical problem of the present
theory.
Nevertheless we expect that polyhomogeneous initial data of the kind
constructed
in \cite{ACh} for which the shear of $\scri$ vanishes will lead to space--times
with metrics which along lightlike directions admit expansions in terms of
$r^{-j}\log^i r $ \cite{footnote},
rather than in
terms of $r^{-j}$ as postulated in \cite{BMS,Sachs} ({\em cf.\ }\cite{ChS}
for a more detailed discussion of that question).

The results presented here immediately lead to the following question:
how much {\em physical} generality does one lose by restricting
oneself to Cauchy data which satisfy the conditions
(\ref{nolog1})--(\ref{nolog2})? These conditions are similar in spirit
to those of Bondi {\em et al.\ } \cite{BMS}, who impose conditions
on the $r^{-2}$ terms in the ``free part of the metric'' at $u=0$ to avoid the
occurrence
of $r^{-j}\log^i r$ terms in the metric at later times.
By doing so, or by imposing (\ref{nolog1})--(\ref{nolog2}),
one gains the luxury of working with smooth conformal completions,
avoiding
all the complications which arise due to the occurrence of log terms
---
but, then, does one overlook some {\em physically significant}
features of radiating gravitating systems?

To obtain a real understanding of
gravitational radiation,
it is therefore
necessary to establish what
asymptotic conditions are appropriate from a physical point of view. The
following are some criteria
which might be considered as physically desirable:
\begin{enumerate}
\item
existence of a well defined notion of total energy;
\label{cond2}
\item
\label{angular}
existence of a well defined notion of angular momentum;
\item \label{cond3}
existence of a development $(
V, \gamma)$ of the initial data set which
admits a ${\scri}$
with a reasonable regularity;
\item\label{cond5} existence of a development of
the data up to $i^+$.
\end{enumerate}
In some situations it might be appropriate to impose only part of
the above conditions.
On the other hand it might perhaps be appropriate to
add to the above
the requirement that the function spaces
considered include those data sets which
arise by evolution from generic initial data which are asymptotically flat
at spatial infinity.

We would
like to emphasize that  {\em it is not known} what regularity conditions
on the conformally compactified metric are {\em necessary} for any of
the above
criteria
to hold.

\acknowledgements
Most of the work presented here was done while P.T.C.\ was visiting
the Centre for Mathematics and its Applications  of the Australian
National University, Canberra;
he was moreover supported in part by the KBN grant \# 2 1047 91 01 and
by the Alexander von Humboldt Foundation.
P.T.C.\ wishes to thank
the Department of
Mathematics of Universit\'e de Tours for friendly hospitality during
part of work on the results presented here; he acknowledges
useful discussions with J.\ Jezierski, G.\ \L ysik, R.\ Mazzeo,
B.\ Schmidt, D.\ Singleton and B.\ Ziemian.
We also wish to thank J.\ Isenberg and an anonymous referee for suggestions
of improvements of a previous version of this letter.


\begin{references}
\bibitem{footnote}{The embedding of the initial
data set into an asymptotically flat  space--time constructed in
\cite{AChWeyl} gives a space--time metric with a Bondi--type expansion which
contains in general
$r^{-1}\log r$ terms. It could
however, happen that there exist better embeddings for which the first $\log$
terms occur with higher powers of $r^{-1}$.}
\end{references}
\end{document}